\documentclass[12pt]{article}
\usepackage{amssymb,amsmath,natbib,graphicx,amsthm,
  setspace,sectsty,anysize,times,dsfont,enumerate}

\usepackage[svgnames]{xcolor}
\usepackage[thinlines]{easytable}

\usepackage{arydshln,relsize,rotating,multirow}
\usepackage{caption}
\usepackage{algorithm}

\newtheorem{theorem}{\sc Theorem}[section]

\marginsize{1.1in}{.9in}{.3in}{1.4in}

\newcommand{\dbl}{\setstretch{1.5}}
\newcommand{\sgl}{\setstretch{1.1}}

\newcommand{\bs}[1]{\boldsymbol{#1}}
\newcommand{\mc}[1]{\mathcal{#1}}
\newcommand{\mr}[1]{\mathrm{#1}}
\newcommand{\bm}[1]{\mathbf{#1}}
\newcommand{\ds}[1]{\mathds{#1}}
\newcommand{\indep}{\perp\!\!\!\perp}

\sectionfont{\noindent\normalfont\large\bf}
\subsectionfont{\noindent\normalfont\normalsize\bf}
\subsubsectionfont{\noindent\normalfont\it}

\usepackage[bottom,hang,flushmargin]{footmisc}

\begin{document}

\sgl 
\pagestyle{empty}

~
\vskip .5cm

\noindent {\LARGE \bf A nonparametric Bayesian analysis of heterogeneous

\vskip .25cm \noindent  treatment effects
in digital experimentation} 

\vskip 1cm
\noindent
Matt Taddy \texttt{(taddy@chicagobooth.edu)}

\noindent
\emph{University of Chicago Booth School of Business}
\setcounter{footnote}{-1}
{\let\thefootnote\relax\footnote{Taddy is also a research fellow at EBay.  The authors thank others at EBay who have contributed, especially Jay Weiler who assisted in data collection.}}

\vskip .25cm
\noindent
Matt Gardner

\noindent
\emph{EBay}

\vskip .25cm
\noindent
Liyun Chen

\noindent
\emph{EBay}

\vskip .25cm
\noindent
David Draper

\noindent
\textit{University of California, Santa Cruz}

\vskip 2cm

\noindent
\textbf{Abstract:} Randomized controlled trials play an important role in how
Internet companies predict the impact of policy decisions and product changes.
In these `digital experiments', different units (people, devices, products)
respond differently to the treatment.  This article presents a fast and
scalable Bayesian nonparametric analysis of such heterogeneous treatment
effects and their measurement in relation to observable covariates.  New
results and algorithms are provided for quantifying the uncertainty associated with
treatment effect measurement via both linear projections and nonlinear
regression trees (CART and Random Forests). For linear projections, our
inference strategy leads to results that are mostly in agreement with those
from the frequentist literature.  We find that linear regression adjustment of treatment effect averages (i.e., post-stratification) can
provide some variance reduction, but that this reduction will be vanishingly
small in the low-signal and large-sample setting of digital experiments. For
regression trees, we provide uncertainty quantification for the machine
learning algorithms that are commonly applied in tree-fitting.   We argue that
practitioners should look to ensembles of trees (forests) rather than individual trees in
their analysis. The ideas are applied on and illustrated through
an example experiment involving 21 million unique users of
\texttt{EBay.com}.

\newpage
\dbl

\pagestyle{plain}
\vskip 1cm
\section{Introduction}

The Internet is host to a massive amount of experimentation.  Online companies
are constantly experimenting with changes to the `user' experience. Randomized
controlled trials are particularly common; they are referred to within
technology companies as `A/B testing' for the random assignment of control
(option A) and treatment (option B) to experimental units (often users, but
also  products, auctions, or other dimensions). The treatments applied can
involve changes to choice of the advertisements a user sees, the flow of
information to users, the algorithms applied in product promotion, the pricing
scheme and market design, or any aspect of website look and function. EBay,
the source of our motivating application, experiments with these
and other parts of the user experience with the goal of making it easier for
buyers and sellers of specific items to find each other.

Heterogeneous treatment effects (HTE) refer to the phenomenon where the
treatment effect for any individual user -- the difference between how they
\emph{would} have responded under treatment rather than control -- is
different from the average.  It is self-evident that HTE
exist: different experimental units (people, products, or devices)  each
have unique responses to treatment.  The task of interest is to measure this
heterogeneity. Suppose  that for each user $i$ with response $y_i$, in either
control or treatment, $d_i = 0$ or $d_i=1$ respectively, there are available
some pre-experiment attributes, $\bm{x}_i$.  These attributes might be related to 
the effect of $d_i$ on $y_i$.  For example, if $y_i$ is 
{\it during-experiment} user spend, then $\bm{x}_i$ could include 
 {\it pre-experiment} spend by user $i$ on the company website. We can then attempt to index the
HTE as a function of $\bm{x}_i$.

Digital (i.e., Internet) A/B experiments differ from  most prior
experimentation in  important ways. First, the sample sizes are enormous.  Our
example EBay experiment (described in Section \ref{sec:data}) has a sample
size of over 21 million unique users. Second, the effect sizes are tiny.  Our
example treatment -- increasing the size of product images  -- has response
standard deviation around 1000 times larger than the estimated treatment
effect.   Finally,  the response of interest (some transaction, such as user
clicks or money spent) tends to be distributed with a majority of mass at
zero, density spikes at other discrete points such as 1 or 99, a long tail,
and variance that is correlated with available covariates. These data features
-- large samples, tiny effects that require careful uncertainty
quantification, and unusual distributions that defy summarization through a
parametric model -- provide a natural setting for nonparametric analysis.

This article proposes a scalable framework for Bayesian nonparametric
analysis of heterogeneous treatment effects.  
Our approach, detailed in
Section \ref{sec:npb}, has two main steps.
\begin{enumerate}[$i$]
\item Choose some \textit{statistic} that is useful for decision making, regardless of the true data distribution.  For example, this could be a difference in means between two groups.
\item Quantify uncertainty about this statistic as induced by the posterior distribution for a flexible Bayesian model of the \textit{data generating process} (DGP). 
\end{enumerate} 
This differs from the usual Bayesian nonparametric analysis strategy, in which a model for the data generating process is applied directly in prediction for future observations.  See \cite{hjort2010bayesian} and \cite{taddy_bayesian_2010} for examples.   In contrast, we consider a scenario where there is a given statistic that will be used in decision making regardless of the true DGP.  The role of Bayesian modeling is solely to quantify uncertainty about this statistic.  For example, in Section \ref{sec:tree} we study regression trees; you do not need to believe that your data was  generated from a tree in order for regression trees to be  useful for prediction. Our statistic in ($i$) is then the output of a tree-fitting algorithm and in ($ii$) we seek to evaluate stability in this algorithm under uncertainty about the true DGP.

We refer to this style of analysis as \textit{distribution-free Bayesian nonparametrics}, in analogy to classical distribution-free statistics \citep[e.g., as in][]{hollander_nonparametric_1999} whose null-hypothesis distribution can be derived under minimal assumptions on the DGP (or without any assumptions at all, as in the case of the rank-sum test of \citealp{wilcoxon1945individual}).  In both Bayesian and classical setups, the statistic of interest is decoupled from assumptions about the DGP.  One  advantage of this strategy in the Bayesian context is that it allows us to apply a class of simple but flexible DGP models whose posterior can be summarized analytically or easily sampled  via a bootstrap algorithm.  That is, we can avoid the computationally intensive Markov chain Monte Carlo algorithms that are usually required in Bayesian nonparametric analysis (and which currently do not scale for data of the sizes encountered in digital experiments).  Moreover,  decoupling  the tool of interest from the DGP model lets us provide guidance to practitioners without requiring them to change how they are processing data and the statistics that they store.  

Our Bayesian DGP model, detailed in Section \ref{sec:npb}, treats the observed
sample as a draw from a multinomial distribution over a large but finite set
of support points.  We place a Dirichlet prior on the probabilities in this
multinomial, and the posterior distribution over possible DGPs is induced by
the posterior on these probabilities.    This Dirichlet-multinomial setup was
introduced in \cite{ferguson_bayesian_1973} as a precursor to his Dirichlet
process model (which has infinite support). We are not the first to propose
its use in data analysis.  Indeed, it is the foundation for the Bayesian bootstrap of \cite{rubin_bayesian_1981}, and \cite{chamberlain_nonparametric_2003} provide a
nice survey of some econometric applications. 
Similarly, our goal is to show the potential for distribution-free Bayesian nonparametric analysis of treatment effects in massive datasets and to understand its implications for some common practices.  

After introducing our motivating dataset in Section \ref{sec:data} and the multinomial-Dirichlet framework in Section
\ref{sec:npb}, the remainder of the paper is devoted to working through the
analysis of two application areas and illustrating the results on our EBay
experiment data. First, Section \ref{sec:linear} considers linear
least-squares HTE projections and their use in adjusting the measurement of
average treatment effects.  We  provide  approximations for the posterior mean
and variance associated with regression-adjusted treatment effects, and the
results predict a phenomena that we have observed repeatedly in practice: the
influence of linear regression adjustment tends to be vanishingly small in the
low-signal and large-sample setting of digital experiments.  Second, Section
\ref{sec:tree} considers the use of the CART \citep{breiman_classification_1984} algorithm for partitioning of units (users) according to their treatment effect heterogeneity.  This is an increasingly popular strategy \citep[e.g.,][]{athey_machine_2015}, and we demonstrate that
there can be considerable uncertainty associated with the resulting partitioning rules.  As a result, we advocate and analyze ensembles of trees that allow one to average over and quantify posterior uncertainty.

\section{Data}
\label{sec:data}

First, some generic notation. For each \textit{independent}  experimental unit $i$,
which we label a `user' following our eBay example, there is a
response $y_i$, a binary treatment indicator $d_i$ with $d_i = 0 \Rightarrow
i \in {\sf c}$ (in control) and $d_i = 1
\Rightarrow i \in {\sf t}$ (in treatment), and a length-$p$ covariate
vector $\bm{x}_i = [x_{i1},\dots,x_{ip}]'$. (We focus on scalar binary
treatment $d_i$ for ease of exposition, but it is straightforward to
generalize our results to multi-factor experiments.) There are
$n_{\sf c}$ users in control, $n_{\sf t}$ in treatment, and $n = (n_{\sf c} +
n_{\sf t})$ in total.  The $n_{\sf d} \times p$ user feature matrices for
control and treatment groups are $\bm{X}_{\sf c}$ and $\bm{X}_{\sf t}$,
respectively, and these are accompanied by response vectors $\bm{y}_{\sf c}$
and $\bm{y}_{\sf t}$. Stacked features and response are $\bm{X} =
\left[\begin{smallmatrix}\bm{X}_{\sf c}\\ \bm{X}_{\sf
t}\end{smallmatrix}\right]$ and $\bm{y} = \left[\begin{smallmatrix}\bm{y}_{\sf
c}\\ \bm{y}_{\sf t}\end{smallmatrix}\right]$, so that ${\sf c} = \{ 1,\dots, n_{\sf c}\}$ are
in control and ${\sf t} = \{ n_{\sf c}+1, \dots, n\}$ are treated.

Our  example experiment involves 21 million users of the website \texttt{EBay.com},
randomly assigned 2/3 in treatment and 1/3 in control over a five week period.
The treatment of interest is a change in image size for items in a user's
\texttt{myEBay} page -- a dashboard that keeps track of items that the user has
marked as interesting. In particular, the pictures are increased from $96\times 96$
pixels in control to $140\times 140$ pixels for the treated.

At EBay, where buyers and sellers transact sales and purchases of items listed
on the website, an important outcome variable is the per-buyer total value of
merchandise bought:  the  amount that a user \emph{spends} on
purchases during the experiment. This $y$  variable is measured in US\$. The response is typical of Internet
transaction data, in that it has
\begin{itemize}
\item a majority at zero, since most users do not make a transaction during the experiment;
\item an long and fat right tail, corresponding to heavy spend by a minority of users;
\item density spikes at, e.g., psychological price thresholds  such as \$1 or \$99; and
\item a variance that
is correlated with both the treatment and sources of treatment
heterogeneity.  For our experiment,  $\mr{sd}(\bm{y}_{\sf t}) =
1153$  is much higher than $\mr{sd}(\bm{y}_{\sf c}) = 970$.
\end{itemize}
Despite these unusual features, raw  spending is \textit{the} business-relevant variable of interest.  For our results to
be useful in decision making we need to understand treatment effects on the
scale upon which EBay makes money.  Common transformations employed by
statisticians to facilitate modeling can lead to unintended consequences.  For
example, focusing on $p ( y > 0 )$ could drive the firm to target low-cost
(and low-profit) items, and  it is not difficult to define scenarios in which
the mean effect  is positive in $\log(y+1)$ but negative in $y$.  This motivates our use of nonparametric methods in analysis of the original  dollar-valued response.

\subsection{User features}

Each user feature vector $\bm{x}_i$, representing potential covariates of
heterogeneity, is constructed from user behavior tracked before the
beginning of the experiment.  Most metrics are aggregated over the four weeks
prior to the start of the experiment, but we also include
  longer-term information in a three-dimensional indicator for whether the user made any purchases in the
past month, quarter, or year.  The metrics tracked include
\begin{itemize}
\item transaction information such as total spending and number of bought items, sold items, or average price per bought item (treated as zero for zero bought items); and 
\item activity information such as counts for site-session visits, page or item views, and actions such as bidding on an item or asking a seller a question. 
\end{itemize}
The variables are tracked in aggregate, as well as broken out by
product category (e.g., collectibles, fashion, or `unknown') and market
platform (e.g., auction or fixed price).  

This gives around 100 total raw variables; they are extremely sparse and
highly correlated.  For the linear analysis of Section \ref{sec:linear}, we
expand the raw data into indicators for whether the variable is greater than
or equal to each of its \emph{positive quintiles}.  That is, there is a binary
$x_{ij}$ element to indicate when  each variable is greater than 0 and when it
is greater than or equal to the $20^{th}$, $40^{th}$, $60^{th}$, and $80^{th}$
percentile of nonzero sample values for that variable.  After collapsing
across equal quintiles (e.g., some variables have up to $60^{th}$ nonzero
percentile equal to one), this results in around 400 covariates.  Finally, we
set $x_{j1} = 1$ unless otherwise specified, so that the regression design
matrices include an intercept.  For the tree partitioning of Section
\ref{sec:tree}, we do no preprocessing and simply input the 100 raw features
to our algorithms.

\section{Bayesian nonparametric sampling model}
\label{sec:npb}

We  employ  Dirichlet-multinomial sampling  as a flexible representation for
the data generating process (DGP).  The approach dates back to \cite{ferguson_bayesian_1973},
\cite{chamberlain_nonparametric_2003} overview it in the context of
econometric problems, and \cite{lancaster_note_2003} and
\cite{poirier_bayesian_2011}  provide detailed analysis of linear projections.
\cite{rubin_bayesian_1981} proposed the Bayesian bootstrap as an algorithm for
sampling from versions of the posterior implied by this strategy, and the
algorithm has since become closely associated with this model.

This model represents the DGP through a  probability mass function on a large
but finite number of possible data points $\bm{z}$ (including response,
covariates, and treatment),  \begin{equation}\label{dgp}
g(\bm{z};~\bs{\theta}) = \frac{1}{|\bs{\theta}|}\sum_{l=1}^L \theta_l \ds{1}{(\bm{z} =\bs{\zeta}_l)}, \end{equation} where $\bs{\mc{Z}} = \{\bs{\zeta}_1 \dots
\bs{\zeta}_L\}$ is the  support of the DGP and $\bs{\theta}$ are
random weights with $\theta_l\geq 0 \;\forall  \;l$. We will often suppress
$\bs{\theta}$ and write $g(\cdot)$ for $g(\cdot;~\bs{\theta})$ unless the
weights need to be made explicit. Here $|\bm{v}|$ denotes $\sum_i
|v_i|$, the $L_1$ norm. Observations are assumed drawn
independently from (\ref{dgp}) by first sampling $l_i$  with
probability $\theta_{l_i}$ and then assigning $\bm{z}_i =
\bs{\zeta}_{l_i}$.   A posterior over $g$ is induced by the posterior over
$\bs{\theta}$.  Functionals of $g$, such as $\ds{E}_gf(\bm{z})$ for arbitrary
function $f$ and where $\ds{E}_g$ implies expectation over $\bm{z}\sim g$, are
thus random variables.

The conjugate prior for the normalized weight vector, $\bs{\theta}/|\bs{\theta}|$,
is a Dirichlet distribution, written
$\mr{Dir}(\bs{\theta}/|\bs{\theta}|; \bs{\nu}) 
\propto \prod_{l=1}^L\left(\theta_l/|\bs{\theta}|\right)^{\nu_l-1}$.  
We  specify a single concentration parameter $a$, such that $\bs{\nu} = [a
\cdots a]'$ for $\ds{E}\left[\theta_l/|\bs{\theta}_l|\right] =  1/L$ and
$\mr{var}(\theta_l/|\bs{\theta}_l|) = (L-1)/[L^2(La+1)]$.  Note that, from the
fact that a Dirichlet realization can be generated as a normalized vector of
independent gamma random variables, this conjugate prior is equivalent to
independent exponential prior distributions on each un-normalized weight:
$\theta_l \stackrel{iid}{\sim} \mr{Exp}(a)$ for $l=1,\dots,L$.

Now, suppose you have the
observed sample $\bm{Z} = [\bm{z}_1
\cdots \bm{z}_n]'$.  For notational convenience, 
we allow $\bs{\zeta}_l=\bs{\zeta}_k$ for $l \neq k$ in the case of repeated
$z_i$ values and write $l_1,\dots, l_n = 1, \dots, n$ so that $\bm{z}_i =
\bs{\zeta}_i$ and $\bm{Z} = [\bs{\zeta}_1 \cdots \bs{\zeta}_n]'$. 
The posterior distribution for $\bs{\theta}/|\bs{\theta}|$  is 
proportional to $\prod_{i=1}^n (\theta_i/|\bs{\theta}|)^{a} \prod_{l=n+1}^L
(\theta_l/|\bs{\theta}|)^{a-1}$.  Again using the constructive definition for the Dirichlet distribution, the  posterior distribution on $\bs{\theta}$  can thus be written as $\theta_l \mid \bm{Z} \stackrel{ind}{\sim} \mr{Exp}\left(a + \ds{1}{(l \leq n)}\right)$.  That is, each weight is an independent exponential random variable with both mean and standard deviation equal to $a + \ds{1}{(l \leq n)}$.

This model places no real restrictions on the DGP beyond that of independence
across observations.  In particular, $L$ could be so large as to include an
practically exhaustive set of values.  However, from here on we will focus on
the limiting prior that arises as $a\rightarrow 0$.  This `non-informative'
limit yields a massive computational convenience: as $a\rightarrow 0$ the
weights for unobserved support points converge to a degenerate random variable
at zero: $p(\theta_l = 0 |\bm{Z}) = 1$ for $l>n$. Our posterior for the DGP is then a multinomial sampling model with random weights on the \emph{observed data points}. We modify notation and
re-write $\bs{\theta} = [\theta_1,
\dots, \theta_n]'$ as the vector of 
weights, so that each DGP is realized in the posterior as
\begin{equation}\label{dgppost}
 g(\bm{z}) \mid \bm{Z} ~=~ \frac{1}{|\bs{\theta}|}\sum_{i=1}^n \theta_i \ds{1}{(\bm{z} = \bm{z}_i)},~~~\theta_i \stackrel{iid}{\sim} \mr{Exp}(1).
 \end{equation}
 The resulting inference model is thus similar to that underlying the frequentist nonparametric bootstrap of \cite{efron1979bootstrap}: the observed support acts as a stand-in for the population support.  The $a \rightarrow 0$ prior is obviously not our `true' prior, but we view this limiting case as a realistic theoretical construction that provides a convenient foundation for uncertainty quantification.

\subsection{Posterior Inference}

Consider some statistic  of the DGP, say $\mathcal{S} \equiv \mathcal{S}(\bs{\theta})$.  This
could be as simple as the DGP mean, written in the posterior as
$\bm{Z}'\bs{\theta}/|\bs{\theta}|$. Following \cite{rubin_bayesian_1981}, we can obtain a sample
from the posterior on $\mathcal{S}$ through a Bayesian bootstrap. 
For
$b=1,\dots, B$:
\begin{itemize}
\item draw $\theta^b_i
\stackrel{iid}{\sim}
\mr{Exp}(1),~i=1,\dots,n$; then 
\item calculate $\mathcal{S}_b =
\mathcal{S}(\bs{\theta}_b)$. 
\end{itemize}

It is also often possible to derive analytic expressions for the posterior on
$\mathcal{S}$.  These are useful for understanding the sources of  uncertainty
and they allow us to avoid Monte Carlo computation.  For example, if $v$ is
some scalar associated with each observation (e.g., an element of $\bm{z}$)
and $\bm{v} = [v_1 \cdots v_n]$ is the sample vector, the DGP mean for $v$ is
written as $\mu_v = \bm{v}'\bs{\theta}/|\bs{\theta}|$.  We can derive the
first two posterior moments for this DGP mean as 
\begin{align}\label{postmoments}
\ds{E}[\mu_v] &= \bar v
\equiv \frac{1}{n}\sum_l v_l \\ 
\mr{var}(\mu_v) &=
\bm{v}'\mr{var}(\bs{\theta}/|\bs{\theta}|)\bm{v} =
\frac{1}{n(n+1)}\bm{v}'\left[\bm{I} - \frac{1}{n}\right]\bm{v} =
\frac{1}{n+1}\left[ \frac{1}{n}\bm{v}'\bm{v}  -  \bar v^2 \right].
\notag
\end{align}
Throughout,  $\ds{E}$ and $\mr{var}$ to refer to posterior mean and variance
operators unless otherwise indicated. In more complex examples, we  proceed
by studying     the posterior on  first-order Taylor series expansions for the
statistic of     interest around the \emph{posterior mean DGP}, which has
$\bs{\theta}= \bm{1}$.  The availability of simple analytic expressions for the posterior mean and variance associated with arbitrary statistics
is an important part of our approach.

\section{Linear HTE projection and adjusted average effects}
\label{sec:linear}

Linear regression is probably the most commonly applied tool in measurement
for heterogeneous treatment effects;  \cite{lin_agnostic_2013} includes a nice
overview.  Specifically, one can study HTE through the difference in ordinary
least-squares (OLS) projections within each treatment group,
\begin{equation}\label{olsdiff} \bm{b}_{\sf t} - \bm{b}_{\sf c} = (\bm{X}_{\sf
t}'\bm{X}_{\sf t})^{-1}\bm{X}_{\sf t}'\bm{y}_{\sf t} - (\bm{X}_{\sf
c}'\bm{X}_{\sf c})^{-1}\bm{X}_{\sf c}'\bm{y}_{\sf c}. \end{equation}   This
statistic is obviously relevant if the response is truly linear in $\bm{x}$.
In that case, under a wide range of additive error models, $\bm{x}'\bm{b}_{\sf d}$
is a consistent and unbiased estimator for the conditional mean
$\ds{E}_f[y\mid \bm{x}, d]$, with $\ds{E}_f$ here denoting the frequentist's
expectation under a true but unknown DGP $f$. In our
setting of randomized controlled trials (i.e., an A/B test), treatment
randomization implies that $d$ is independent from both $y$ and $\bm{x}$, so
that $\bm{x}'(\bm{b}_{\sf t} - \bm{b}_{\sf c}) \approx \ds{E}_f[y\mid \bm{x}, d=1] -
\ds{E}_f[y\mid \bm{x}, d=0]$ can be interpreted as the conditional average
\textit{causal} effect of $d$ on $y$ given $\bm{x}$ (see
\citealp{imbens_nonparametric_2004} and \citealp{imbens2015causal} for background on causal inference).
Moreover, there are many applications where the difference in linear
projections will be useful  even if the true DGP mean is nonlinear in
$\bm{x}$.  For example, we consider in Section \ref{sec:linear}.2 below the
common frequentist practice of using  $\bm{\bar x}'(\bm{b}_{\sf t} - \bm{b}_{\sf c})$  as a measure of the  average
treatment effect.

\subsection{Linear analysis of heterogeneous treatment effects}
 
From our nonparametric Bayesian perspective, the difference between treatment
group {\it population} OLS projections is a statistic of the underlying DGP.
Write the group projections as \begin{equation}\label{popols} \bs{\beta}_{\sf d} =
(\bm{X}_{\sf d}'\bs{\Theta}_{\sf d}\bm{X}_{\sf d})^{-1}  \bm{X}_{\sf
d}'\bs{\Theta}_{\sf d}\bm{y}_{\sf d}, \end{equation}  where $\bs{\Theta}_{\sf
d} = \mr{Diag}(\bs{\theta}_{\sf d})$ and $\bs{\theta}_{\sf d}$ is the 
sub-vector of weights for $i \in {\sf d}$ (recall $d_i =1 \Leftrightarrow i \in
{\sf t}$ and $d_i =0 \Leftrightarrow i \in {\sf c}$). Since the treatment
groups are independent, $\bs{\theta}_{\sf t} \indep \bs{\theta}_{\sf c}$ and
$\bs{\beta}_{\sf t} \indep \bs{\beta}_{\sf c}$  so that the posterior for the
difference $ \bs{\beta}_{\sf t} - \bs{\beta}_{\sf c} $  can be derived directly from
the posterior on (\ref{popols}).  A Bayesian bootstrap for  $\bs{\beta}_{\sf t} -
\bs{\beta}_{\sf c}$ thus takes the difference between \textit{weighted-OLS} fits
for treatment and control groups, with weights drawn independently from the
$\mr{Exp}(1)$ distribution.

To provide an analytic summary of the posterior on $\bs{\beta}_{\sf t} -
\bs{\beta}_{\sf c}$, we  proceed by  studying the
\textit{exact} posterior on first-order Taylor series expansions for $\bs{\beta}_{\sf d}$ around its value at the posterior mean DGP, where
$\bs{\theta}= \bm{1}$.   \citet{lancaster_note_2003} and
\citet{poirier_bayesian_2011} also apply this approach, under the same Dirichlet-multinomial model that we use, in analysis of OLS projections.  
A first-order Taylor approximation to the group-specific population OLS 
in (\ref{popols}) is
\begin{equation}\label{taylorols}
\bs{\tilde \beta}_{\sf d} = \bs{\hat \beta}_{\sf d} + 
\nabla \bs{\beta}_{\sf d}\big |_{\bs{\theta}_{\sf d}=\bm{1}} (\bs{\theta}_{\sf d} - \bm{1}),
\end{equation}
where $\bs{\hat \beta}_{\sf d} = \bs{\beta}_{\sf d}|_{\bs{\theta}_{\sf d}=\bm{1}} = \left(\bm{X}_{\sf d}'\bm{X}_{\sf d}\right)^{-1}\bm{X}_{\sf d}'y_{\sf d}$ is the OLS projection at posterior mean DGP.  As detailed in Appendix \ref{sec:olsgrad}, the $p\times n$ gradient  is 
$
\nabla \bs{\beta}_{\sf d}
 =  
(\bm{X}_{\sf d}'\bs{\Theta}_{\sf d}\bm{X}_{\sf d})^{-1} \bm{X}_{\sf d}\mr{diag}(\bm{y}_{\sf d}-\bm{X}_{\sf d}\bs{\beta}_{\sf d})
$
and 
\begin{equation}\label{olsgrad}
\nabla \bs{\beta}_{\sf d}|_{\bs{\theta}_{\sf
d}=\bm{1}} = (\bm{X}_{\sf d}'\bm{X}_{\sf d})^{-1} \bm{X}_{\sf
d}\bm{R}_{\sf d}
\end{equation} where $\bm{R}_{\sf d} = \mr{diag}(\bm{r}_{\sf d})$  and $\bm{r}_{\sf d}$ is the vector of 
group-specific OLS residuals, $r_{{\sf d} i} = y_i - \bm{x}_i'\bs{\hat\beta}_{\sf d}$ for $i\in {\sf d}$.  Since each $\theta_i$ has an independent exponential posterior distribution, such that $\mr{var}(\bs{\theta}_{\sf d})=\bm{I}_{n_{\sf d}}$, the approximation in (\ref{popols}) has \emph{exact} posterior variance
\begin{equation}\label{olsvar}
\mr{var}(\bs{\tilde \beta}_{\sf d}) = \left(\nabla \bs{\beta}_{\sf d}|_{\bs{\theta}_{\sf
d}=\bm{1}}\right) \mr{var}(\bs{\theta}_{\sf d})\left(\nabla \bs{\beta}_{\sf d}|_{\bs{\theta}_{\sf
d}=\bm{1}}\right)' = (\bm{X}^{\prime}_{\sf d}\bm{X}_{\sf d})^{-1}\bm{X}_{\sf d}^{\prime}\bm{R}_d\bm{R}_d\bm{X}_{\sf d}(\bm{X}_{\sf d}^{\prime}\bm{X}_{\sf d})^{-1}.
\end{equation}
This matches the Huber-White heteroskedastic-consistent
OLS variance formula
\citep{white_heteroskedasticity-consistent_1980}; hence, the first-order approximate nonparametric Bayesian posterior
variance formula is the same as a widely used frequentist estimator for sampling variance in OLS.

\subsubsection{Example:  treatment effect heterogeneity between new and old users}

\begin{figure}
\includegraphics[width=\textwidth]{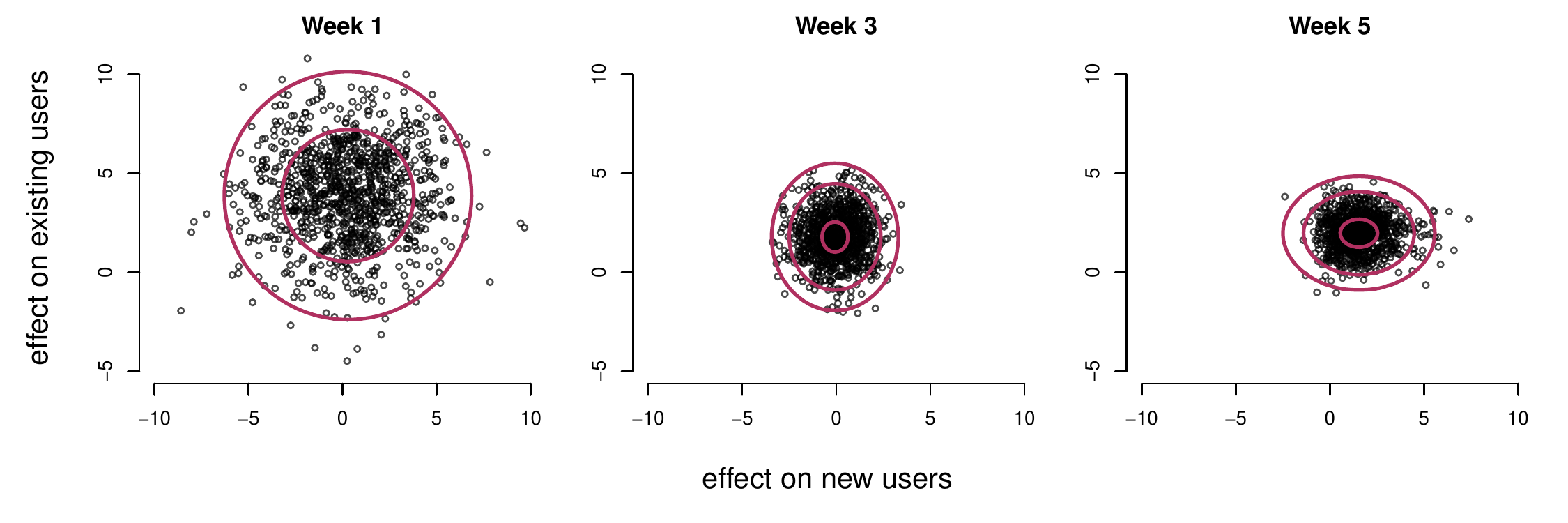}
\vskip -.25cm
\caption{\label{fig:gammapost} Conditional average treatment effects for new vs existing users.  Posterior samples of $\bs{\beta}_{\sf t} -
\bs{\beta}_{\sf c}$ given data accumulated through 1, 3, and 5 experiment weeks.   The contours correspond to values of $\{0.1,0.01,0.001\}$ in a bivariate normal density centered at  $\bm{b}_{\sf t} - \bm{b}_{\sf c}$ and with variance $\mr{var}(\bs{\tilde\beta}_{{\sf t}}) +
\mr{var}(\bs{\tilde\beta}_{{\sf c}})$. }
\end{figure}

To briefly illustrate these ideas, we turn to the EBay experiment data of
Section \ref{sec:data} and consider the change in conditional average
treatment effect across a single influential covariate:  whether or not the
user has made a purchase in the past year.  Any user who has made such a
purchase is labeled \textit{existing}, while those who have not are
\textit{new}.

Write the feature vector as $\bm{x}_i = [x_{i\,\textit{new}},x_{i\,\textit{exist}}]'$, with each
element an indicator for whether the user is \textit{new} or \textit{existing} (so
that $x_{i\,\textit{new}}x_{i\,\textit{exist}}=0$ always).  In this simple case,  the four random
variables of interest -- $\beta_{{\sf t}\,\textit{new}},\beta_{{\sf
t}\,\textit{exist}},\beta_{{\sf c}\,\textit{new}},\beta_{{\sf c}\,\textit{exist}}$ -- each correspond to the
means associated with independent sub-vectors of $\bs{\theta}$ and are thus independent from each other.  We can apply the
formulas in (\ref{postmoments}) to obtain the exact posterior moments for each and hence for the treatment
effects,
$\beta_{{\sf t}j} - \beta_{{\sf c}j}.$ 
Working through the formula in (\ref{olsvar}) shows that, in this setup, the
variance of our first-order approximation is very nearly equal to the true
posterior variance: 
\begin{equation} \label{betavar}
\ds{E}[\bs{\beta}_{\sf d}] = \bm{b}_{\sf
d} ~~~~~\text{and}~~~~~
\mr{var}(\beta_{{\sf d}j}) = \frac{n_{ {\sf d}j }}{n_{{\sf d}j}+1}\mr{var}(\tilde
\beta_{{\sf d}j}),
\end{equation}
where $n_{{\sf d}j} = \sum_{i \in \sf d} x_{ij}$ is 
the number of members of treatment group ${\sf d}$ that were of user type $j$.
Note that the variances in (\ref{betavar}) would be exactly equal if the series in
(\ref{taylorols}) was expanded for $\bs{\theta}/|\bs{\theta}|$ around $\bm{1}/n$, as is done in
\cite{poirier_bayesian_2011}, instead of for $\bs{\theta}$ as we have here.

Figure \ref{fig:gammapost} presents the posterior distribution for
$\bs{\beta}_{\sf t} - \bs{\beta}_{\sf c}$. Results are shown
conditional upon the 7.45 million users who  visited their \texttt{myEBay} page in the
first week of the experiment, the 10.97 million users who visited in the first
three weeks, and the 13.22 million who visited at any time during the
experiment. The posteriors are summarized through 1000 draws from the Bayesian
bootstrap  and as well as through a normal distribution centered at the sample
OLS difference, $\ds{E}[\bs{\beta}_{\sf t} - \bs{\beta}_{\sf c}] = \bm{b}_{\sf
t} - \bm{b}_{\sf c}$,  and with variance $\mr{var}(\bs{\beta}_{\sf t} -
\bs{\beta}_{\sf c}) \approx \mr{var}(\bs{\tilde\beta}_{{\sf t}}) +
\mr{var}(\bs{\tilde\beta}_{{\sf c}})$. Following the discussion above,  this approximate
variance is very nearly equal to the true posterior variance and the \textit{new}
and \textit{existing} user average treatment effects are independent from each
other in the posterior. Comparing the normal  approximation to the
bootstrap samples, we see that the former's tails are thinner than those of
the true posterior. Even after 5 weeks and 13 million exposed users, the posterior is
skewed with a long right tail for $\beta_{{\sf t}\,\textit{new}}-\beta_{{\sf c}\,\textit{new}}$.

\subsection{Average treatment effects}

A primary goal of many A/B experiments is to infer the \textit{average treatment effect} (ATE): the average difference between what a user will spend if they experience option B versus if they experience option A.  This value, multiplied by the number of users on your website, is the change in revenue that you will receive if moving from option A to option B.  

Again referring the reader to \cite{imbens2015causal} for additional detail,
the  massive benefit of randomized controlled trials is that each user's group
assignment, $d$, is independent of their treatment effect.  The difference
between the expected response given that a user has been assigned to treatment
versus control can then be interpreted causally as the treatment effect.  This
motivates the usual measure of ATE for a randomized trial as the difference in
group means, $\bar y_{\sf t} - \bar y_{\sf c}$.  Writing $\mu_{\sf d} =
\bm{y}_{\sf d}'\bs{\theta}_{\sf d}/|\bs{\theta}_{\sf d}|$, the corresponding
DGP statistic of interest is $\mu_{\sf t} - \mu_{\sf c}$.  Applying
(\ref{postmoments}), we get the posterior moments
\begin{equation}\label{margvar} \ds{E}[\mu_{\sf t} - \mu_{\sf c}] = \bar
y_{\sf t} - \bar y_{\sf c}~~~~\text{and}~~~~\mr{var}(\mu_{\sf t} - \mu_{\sf
c}) =  \frac{1}{n_{\sf t}(n_{\sf t} +1)}s^2_{\bm{y}_{\sf t}}+ \frac{1}{n_{\sf
c}(n_{\sf c} +1)}s^2_{\bm{y}_{\sf c}} \end{equation} with $s^2_{\bm{v}} =
\bm{v}'\bm{v} - n_v \bar{\bm{v}}^2$  the sum-squared-error for generic
length-$n_v$ vector $\bm{v}$.  The posterior is thus centered on the usual
estimate for the difference in means between two populations, with variance
that is a slight deflation of the common sampling variance formula for this
estimator.

There is a large frequentist literature on inference for ATE that advocates
replacing $\bar y_{\sf t} - \bar y_{\sf c}$ with alternative metrics that
condition upon the information in covariates, $\bm{x}$.  One recently popular
strategy promotes a {\it regression-adjusted average treatment effect}
\citep[e.g.,][]{lin_agnostic_2013},
\begin{equation}\label{raate}   \bm{\bar
x}'(\bm{b}_{\sf t} - \bm{b}_{\sf c}). 
\end{equation}  
 The advantage of
such estimators is that they can have lower sampling variance than $\bar y_{\sf t} -
\bar y_{\sf c}$.  When (\ref{raate}) is also
unbiased for some definition of the \textit{true} ATE \citep[see][for a survey
of frequentist inferential targets]{imbens_nonparametric_2004}, it thus leads
to more efficient estimation.

One way to justify (\ref{raate}) as representative of the ATE is to assume a
linear relationship between $\bm{x}$ and $y$ conditional upon $d$. Linearity
holds trivially if $\bm{x}$ represents a partitioning of the data into mutually exclusive strata.   Application of (\ref{raate}) in this situation is referred to
as {\it post stratification}; see \cite{deng_improving_2013} for use in the
context of digital experiments and \citet{miratrix_adjusting_2013} for a
detailed theoretical overview. But (\ref{raate}) can also be justified for
situations where linearity does not hold. \citet{berk_covariance_2013},
\citet{pitkin_improved_2013}, \citet{lin_agnostic_2013}, and
\citet{lin_comments_2014} study ATE estimation under a variety of assumed
sampling models.  In each case, they show that $\bm{\bar
x}'(\bm{b}_{\sf t} - \bm{b}_{\sf c})$ is an
asymptotically more efficient estimator of average treatment effects than the
simple difference in means, $\bar y_{\sf t} - \bar y_{\sf c}$.  

Both $\bar y_{\sf t} - \bar y_{\sf c}$ and (under the sampling models  in the
literature cited above) $\bm{\bar x}'(\bm{b}_{\sf t} - \bm{b}_{\sf c})$ are
unbiased for commonly targeted definitions of the  ATE (e.g., the sample
average treatment effect -- what would have been the average effect if we
could have observed every user under \textit{both} treatment and control). The
argument for use of the latter regression-adjusted statistic is  that it
reduces your estimation variance. Quantifying the amount of variance reduction
requires  assumptions about the underlying DGP (see the Berk,
Pitken et al. and Lin references above)  and it is not always clear what one
should expect in real examples.  Our personal experience with digital
experiments is that the frequentist nonparametric bootstrap often shows
little, if any, variance reduction after regression adjustment.

Consider covariate averages projected through the population
OLS  lines from (\ref{popols}),  \begin{equation}\label{rateg}
\bs{\mu}_{\bm{x}}'(\bs{\beta}_{\sf t} - \bs{\beta}_{\sf c}) =
\frac{1}{|\bs{\theta}|}\bs{\theta}'\bm{X}\bigg( (\bm{X}_{\sf
t}'\bs{\Theta}_{\sf t}\bm{X}_{\sf t})^{-1}  \bm{X}_{\sf t}'\bs{\Theta}_{\sf
t}\bm{y}_{\sf t} - (\bm{X}_{\sf c}'\bs{\Theta}_{\sf c}\bm{X}_{\sf c})^{-1}
\bm{X}_{\sf c}'\bs{\Theta}_{\sf c}\bm{y}_{\sf c}\bigg), \end{equation}  where
$\bs{\mu}_{\bm{x}} = \bm{X}'\bs{\theta}/|\bs{\theta}|$. This is a
random variable; it is the Bayesian nonparametric target analogous to
$\bm{\bar x}'(\bm{b}_{\sf t} - \bm{b}_{\sf c})$.
As detailed in Appendix \ref{sec:regappendix}, a first-order approximation to 
(\ref{rateg}) is  $\bm{\bar
x}'\left(\bs{\tilde\beta}_{\sf t} -
\bs{\tilde \beta}_{\sf c}\right)$.  This has exact posterior mean
 $\bm{\bar x}'(\bm{b}_{\sf t} - \bm{b}_{\sf c})$ and 
 Theorem \ref{ratevarthm} shows that \begin{equation}\label{roughatevar}
\mr{var}\!\left[\bm{\bar
x}'\left(\bs{\tilde\beta}_{\sf t} -
\bs{\tilde \beta}_{\sf c}\right)\right] \approx
\mr{var}(\bar y_{\sf t} - \bar y_{\sf c}) - \left( \frac{R^2_{\sf t}s^2_{\bm{y}_{\sf
t}}}{n^2_{\sf t}} +   \frac{R^2_{\sf c}s^2_{\bm{y}_{\sf c}}}{n^2_{\sf c}}
\right), 
\end{equation} where  $R^2_{\sf d} = 1 - s^2_{\bm{r}_{\sf
d}}/s^2_{\bm{y}_{\sf d}}$ is the proportion of
deviance explained for the group-${\sf d}$ OLS regression. Thus regression
adjustment yields a large variance reduction only if elements of $\bm{x}$ have
large covariances with $y$ and if the sample size is not too big. Since our digital experiments have  tiny signals and massive samples,  we should expect precisely what we have experienced in practice: little variance reduction from regression adjustment.

\subsubsection{Example: regression-adjustment for average treatment effects}

Table
\ref{tab:perweek} shows  posterior inference for average treatment effects in our EBay
experiment, conditional upon the data accumulated through each week of the
experiment. We have applied here the full set of user features, expanded into
length-$400$ covariate vectors $\bm{x}_i$ as described in  Section
\ref{sec:data}. The table shows posterior means and standard deviations for
three ATE statistics: the unadjusted difference in group means, $\mu_{\sf t} -
\mu_{\sf c}$; the regression-adjusted ATE, $\bs{\mu}_{\bm{x}}'(\bs{\beta}_{\sf
t} - \bs{\beta}_{\sf c})$; and the approximation to this  regression-adjusted
ATE, $\bm{\bar{x}}'(\bs{\tilde\beta}_{\sf t} - \bs{\tilde\beta}_{\sf c})$.
For later reference, we also include the average difference between regression trees fit in each treatment group, labeled $\ds{E}_{g}[{\hat y}_{\sf t}(\bm{x}) - {\hat y}_{\sf c}(\bm{x})]$; we defer discussion to Section \ref{sec:tree}.
Posterior means and standard deviations for $\bs{\mu}_{\bm{x}}'(\bs{\beta}_{\sf
t} - \bs{\beta}_{\sf c})$ are obtained via Bayesian bootstrap sampling, while
moments for the other statistics area available analytically as described
above.

As predicted by Theorem \ref{ratevarthm}, since our samples are large and our correlations are small, there is practically zero reduction
in variance between $\mu_{\sf t} - \mu_{\sf c}$ and
$\bm{\bar{x}}'(\bs{\tilde\beta}_{\sf t} - \bs{\tilde\beta}_{\sf c})$.  The
standard deviations are the same up to two decimal places (\$0.01) in every case.
Moreover, the Bayesian bootstrap posterior sample for $\bs{\mu}_{\bm{x}}'(\bs{\beta}_{\sf
t} - \bs{\beta}_{\sf c})$ has means and standard deviations that are similar to those of its approximation, $\bm{\bar{x}}'(\bs{\tilde\beta}_{\sf t} - \bs{\tilde\beta}_{\sf c})$, giving evidence to support the practical relevance of (\ref{roughatevar}).  While neither this example nor our analytic results imply that regression-adjustment cannot yield lower-variance inference, it does fit with our practical experience that such adjustments makes little difference in experiments like the one studied here.

\begin{table}[bth]
\small\vskip .5cm
{\it \normalsize Posterior Mean (and SD)}
\begin{center}\dbl
\vskip -.5cm
\begin{tabular}{c|ccccc}
 \multicolumn{1}{c}{}& week 1 & week 2 & week 3 & week 4 & week 5 \\
  \hline
$\mu_{\sf t} - \mu_{\sf c}$   & 3.30 (2.04) & 1.34 (1.24) & 1.44 (0.95) & 1.71 (0.80) & 1.90 (0.76) \\
$\bs{\mu}_{\bm{x}}'(\bs{\beta}_{\sf t} - \bs{\beta}_{\sf c})$ & 2.95 (2.07) & 1.26 (1.29) & 1.39 (0.84) & 1.55 (0.72) & 1.83 (0.75) \\
$\bm{\bar{x}}'(\bs{\tilde\beta}_{\sf t} - \bs{\tilde\beta}_{\sf c})$  & 3.05 (2.04) & 1.15 (1.24) & 1.29 (0.95) & 1.59 (0.80) & 1.80 (0.76)  \\
${\hat y}_{\sf t} - {\hat y}_{\sf c}$ & 2.99 (1.46) & 1.03 (1.46) & 1.20 (0.90) & 1.74 (0.80) & 1.75 (0.82) \\
 \hline
 \multicolumn{1}{c}{number of users, in mil} & 7.45~~~  & 9.48~~~ & 10.97~~ & 12.20~~ & 13.22~~
\end{tabular}
\end{center}
\vskip -.25cm
\caption{\label{tab:perweek} Posterior means (and standard deviations) for ATE statistics conditional upon the sets of users who visited their
\texttt{myEBay} page through 1-5 weeks, cumulative.  Values for
$\mu_{\sf t} - \mu_{\sf c}$ and $\bm{\bar{x}}'(\bs{\tilde\beta}_{\sf t} - \bs{\tilde\beta}_{\sf c})$ are exact, while  moments for 
$\bs{\mu}_{\bm{x}}'(\bs{\beta}_{\sf t} - \bs{\beta}_{\sf c})$ are based on 100 draws from the Bayesian bootstrap. Finally, ${\hat y}_{\sf t} - {\hat y}_{\sf c}$ refers to the difference between treatment-group-specific regression trees, as detailed in Section \ref{sec:tree}.2; posterior moments here are based on two forests of 1000 trees each.}
\end{table}

\section{Regression tree prediction for HTE}
\label{sec:tree}

Regression trees partition the feature (covariate) space into regions of
response homogeneity, such that the response associated with any point in a
given partition can be predicted from the average for that of its neighbors.
The partitions are typically formed through a series of binary splits, and
after this series of splits each terminal {\it leaf} node contains a
rectangular subset of the covariate support. Advantages of using trees as
prediction rules include that they can model a response that is nonlinear in
the original covariates, they can represent complex interactions (every
variable that is split upon is interacting with those above and below it in
the tree), and they allow for error variance that changes with the covariates
(there is no homoscedasticity restriction across leaves).    Through their
implementation as part of Random Forest
\citep{breiman_random_2001} or gradient boosting machine
\citep{friedman2001greedy} ensembles, it is difficult to overstate the extent
to which trees play a central role in contemporary industrial machine
learning.

The CART algorithm of \cite{breiman_classification_1984} is the most common
and successful recipe for building trees.  It grows greedily and recursively:
for a given node (subset of data), a split location is chosen to minimize some
impurity (sum-squared-error for regression trees)  across the two resulting
\textit{children}; this splitting procedure is repeated on each child, and
hence recursively until the algorithm encounters a stopping rule (e.g., if a
new child contains fewer observations than a specified minimum leaf size).
After the tree is fit, it is common to use cross-validation to {\it prune} it
by evaluating whether the splits near to the leaves improve out-of-sample prediction and
removing those that do not.  Variations on CART include the random removal of
input dimensions as candidates for the split location at each impurity
minimization.  The resulting prediction rule is then the average across
repeated runs of this randomized-input CART  \citep[e.g.,
see][]{breiman_random_2001}. Introduction of such stochasticity can improve
upon the performance of greedy search in datasets where, e.g., you have high-dimensional inputs.

To quantify uncertainty for CART, we study a {\it population} CART algorithm that (analogously to the population OLS in (\ref{popols})) optimizes over a realization of our Bayesian nonparametric DGP model from (\ref{dgppost}). Consider a node $\eta$, containing a subset of the data indices $1,\ldots,n$. This node is to be partitioned into two child nodes according to a binary split on one of the covariate locations:  a split on  input $j$ of observation $k$, say $x_{kj} = x$, so that the two resulting child nodes are $\textit{left}(\eta,j,x) = \{i : x_{ij} \leq x, i\in \eta\}$ and $\textit{right}(\eta,j,x) = \{i : x_{ij} > x, i \in \eta\}$.  Given a realization of the DGP weights $\bs{\theta}$, the population CART algorithm chooses $x_{kj}$ to minimize 
\begin{equation}\label{impurity}
\mc{E}_{\textit{left}(\eta,j,x)}(\bs{\theta}) + \mc{E}_{\textit{right}(\eta,j,x)}(\bs{\theta}),
\end{equation}
where for a generic node $\textsf{s} \subseteq \{1,\ldots,n\}$ the impurity (error) is 
\begin{equation}\label{sse}
\mc{E}_\textsf{s}(\bs{\theta})  = \sum_{i \in \textsf{s}} \theta_i (y_i - \mu_\textsf{s})^2 ~~~~\text{with}~~~~\mu_\textsf{s} = \bm{y}_\textsf{s}'\bs{\theta}_\textsf{s}/|\bs{\theta}_\textsf{s}|.
\end{equation}
As in sample CART, this splitting is repeated recursively until we encounter a
stopping rule. For randomized-input versions of CART, one minimizes the same
DGP-dependent impurity in (\ref{impurity}) but over a random subset of
candidate split dimensions $J \subset \{1,\ldots,p\}$.  The \textit{statistic} of interest is then itself a random object: we have decoupled variability due to uncertainty about the DGP  from algorithmic stochasticity that does not diminish as you accumulate data.

The posterior over trees (i.e., over CART fits) can be sampled via the Bayesian bootstrap of Section \ref{sec:npb}.1.  This leads to a posterior sample of trees that we label a \textit{Bayesian Forest}.  The algorithm is studied in detail in \cite{taddy_forests_2015}, along with an Empirical Bayes approximation  for computation in distribution across many machines.  \cite{taddy_forests_2015} demonstrate that the average prediction from a Bayesian Forest (i.e., the posterior mean) outperforms prediction from a  single CART tree (with cross-validated pruning) and many other common tree-based prediction algorithms.  Of particular interest,  Bayesian Forests tend to perform similarly to, although usually slightly better than, Random Forests.  The only difference between the  two algorithms is that while the Bayesian Forest uses independent $\mr{Exp}(1)$ observation weights, the Random Forest draws a vector of discrete weights from a multinomial distribution with probability $\bm{1}_n/n$ and size $n$ (this is the frequentist nonparametric bootstrap).

\subsection{Single tree prediction for HTE}

Bayesian Forests provide
uncertainty quantification for a machine learning algorithm -- CART -- that is
often viewed as a black-box.  
Recently published applications of regression trees in HTE prediction include
\citet[][]{foster2011subgroup} for medical clinical trials and
\citet[][]{dudik2011doubly} for  user browsing behavior, and  we have observed that CART is commonly employed in industry for the segmentation
of customers according to their response to advertisement and promotions.

\citet{athey_machine_2015} study various strategies 
for the use of CART-like algorithms in prediction of HTE.  We will focus on
their \textit{transformed outcome tree} (TOT) method, which is simply the
application of CART in prediction of a transformed response, $y^\star$, which
has expected value equal to the treatment effect of interest. In the language
of the Neyman-Rubin causal model \citep[e.g.,][]{rubin2011causal}, each unit
of observation $i$ in an experiment is associated with two \textit{potential
outcomes}: $\upsilon_i(0)$, their response if they are
allocated to the control group; and $\upsilon_i(1)$, their response under
treatment.  Of course, only one of these two potential outcomes is ever realized and observed:
$y_i = \upsilon_i(d_i)$. \citet{athey_machine_2015} define 
\begin{equation}\label{ystar}
y^\star_i = y_i \frac{d_i - q}{q(1-q)},
\end{equation}
where $q$ is the probability of treatment ($q=2/3$ in our EBay example).
Then, with $\ds{E}_{d}$ denoting expectation over unknown independent treatment allocation $d_i$,
\begin{equation}\label{eystar}
\ds{E}_{d}[y^\star_i | \bs{\upsilon}_i] = q \upsilon_i({\sf t})\frac{1 - q}{q(1-q)}
- (1-q) \upsilon_i({\sf c})\frac{q}{q(1-q)}
= \upsilon_i({\sf t})- \upsilon_i({\sf c}).
\end{equation}
 Thus a tree that is trained to fit the expectation for $y^\star_i$ can be used to predict the treatment effect.

\begin{figure}[p]
\textbf{~~~1 Week \hskip 6cm 5 Weeks}\\
\includegraphics[width=.38\textwidth]{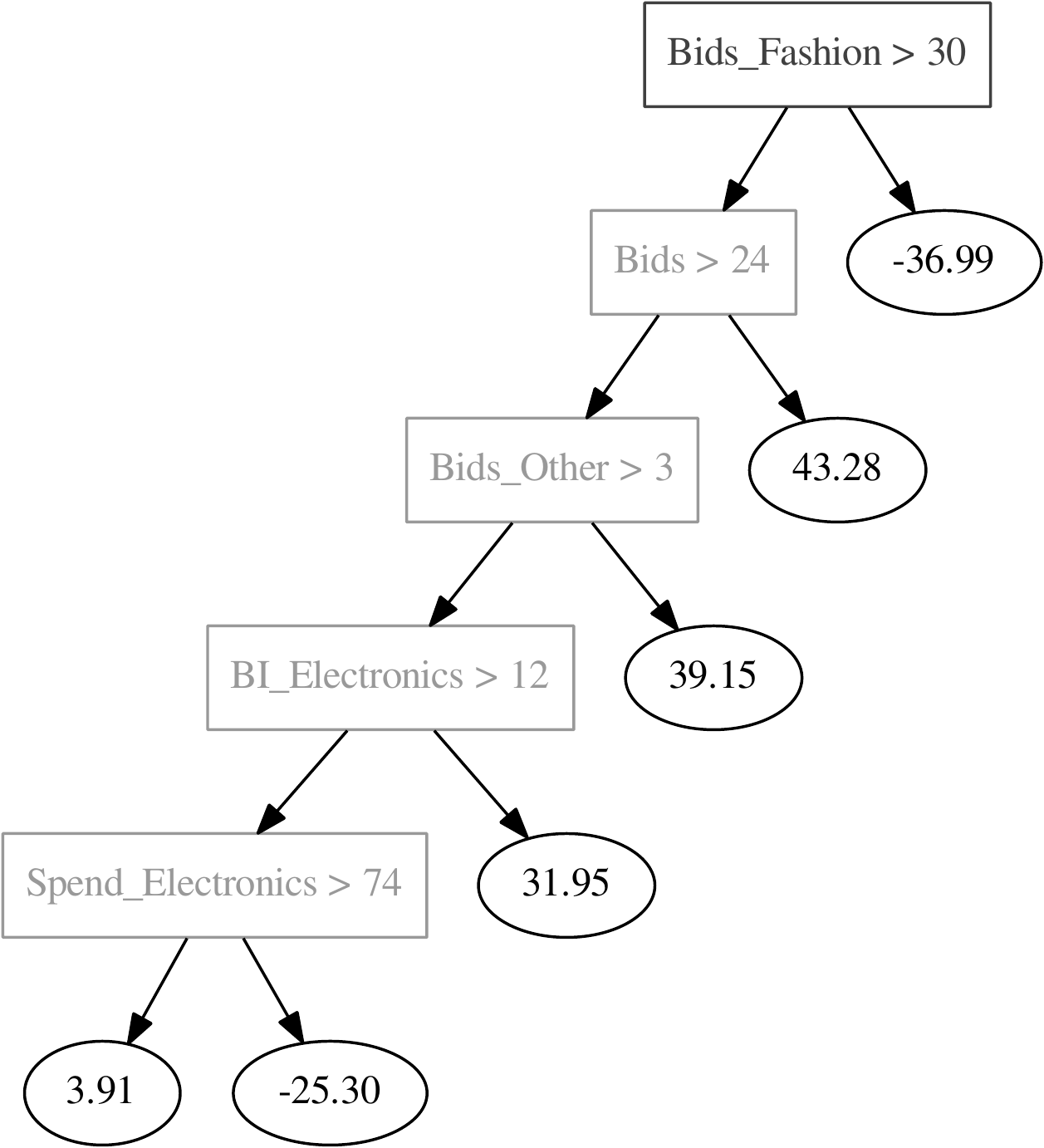}\!
\includegraphics[width=.62\textwidth]{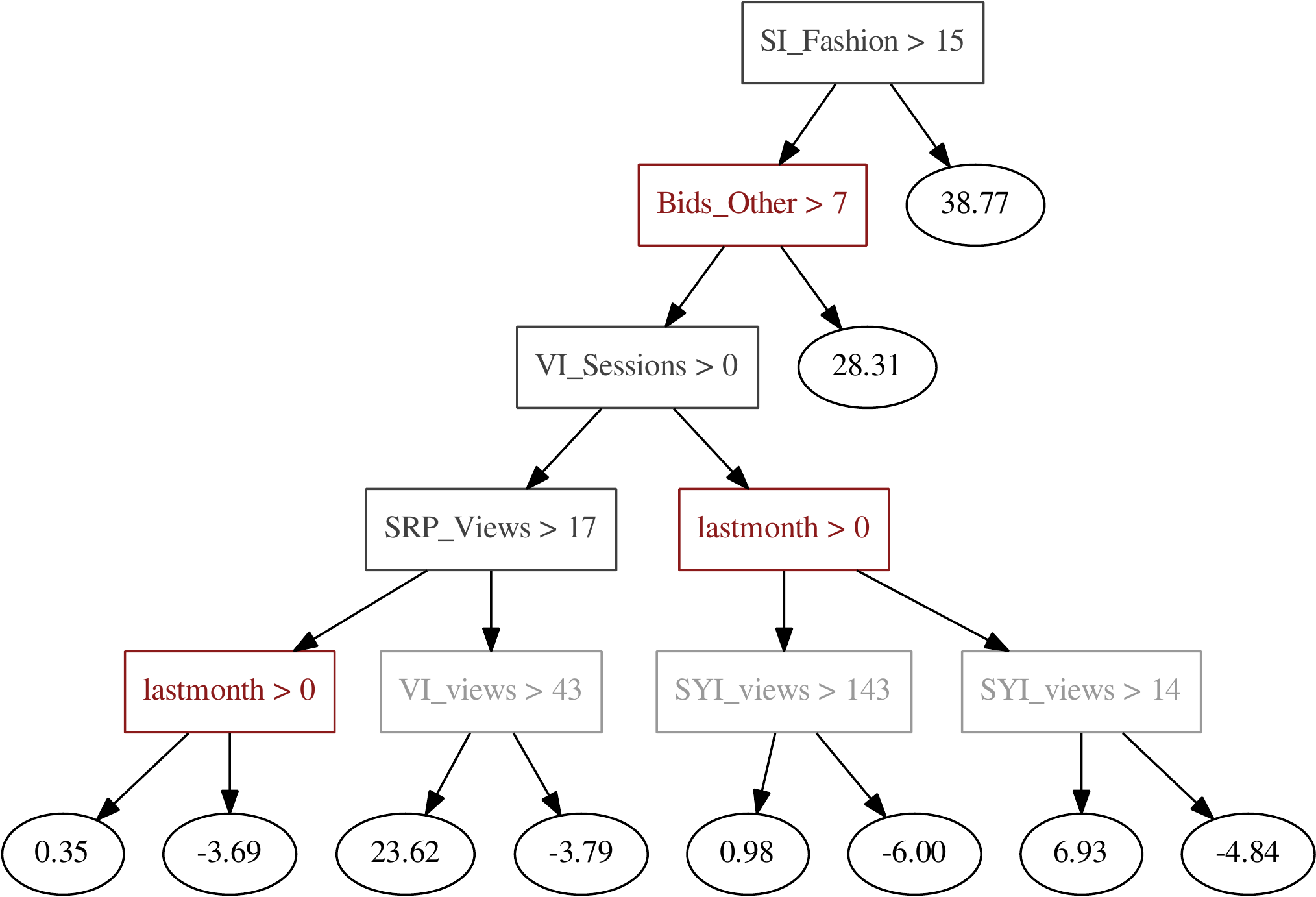}

\vskip .25cm
\caption{\label{fig:trees} Sample TOT (CART for $y^\star$) trees after one and five weeks of
experimentation.  The  algorithm was run with a minimum leaf size of 100,000
users and a maximum depth of 5 internal (splitting) nodes.  The right-hand
children contain data for which the split condition is true, and left-hand
children are the complement.  Leaves are marked with the predicted treatment effect
in that leaf. Decision nodes are colored for the posterior
probability (see Table \ref{tab:trees}) that the corresponding variable is
included in the TOT fit for new a DGP realization: {\color{black!50} light
grey for $\mr{p} <\tfrac{1}{3}$}, {\color{black!75} dark grey for $\mr{p} \in
\left[\tfrac{1}{3},\tfrac{1}{2}\right)$}, and {\color{DarkRed} red for $\mr{p}
\geq \tfrac{1}{2}$}. \\ ~\\
Terminology: \textit{Views} is the count of page
impressions, \textit{Bids} is the count of bids on the auction platform, \textit{Sessions} is the number of sessions where an event
occurs, \textit{SRP} is a search results page, \textit{VI} is where the user
clicked on a item to view details, \textit{BI} is the count of bought items,
  \textit{SI} denotes the count
of items sold by the user and \textit{SYI} denotes the count of items that the
user attempted to sell. 
\textit{Other}, \textit{Fashion}, and \textit{Electronics} denote restriction of the associated metrics to activity in these product categories. Finally, \textit{lastmonth} indicates whether the user made any purchases in the month prior to the experiment.}

\vskip 1cm
{\small
\begin{tabular}{l|ccccc}
\multicolumn{6}{l}{\it Variable split probabilities}\\[.4cm]
\textbf{1 Week}& \multicolumn{5}{c}{\textit{depth in tree}}\\
 & $1$ & $\leq 2$ & $\leq 3$ & $\leq 4$ & $\leq 5$ \\
\hline
\rule{0pt}{1.25\normalbaselineskip}\textit{BidsFashion} & 
.32 & .35 & .38 & .39 & .40 \\
\textit{Bids} & .02 & .08 & .14 & .23 & .27 \\
\textit{Bids\,Other} & .15 & .20 & .22 & .23 & .23 \\
\textit{BI\,Electronics} & .02 & .04 & .11 & .19 & .22 \\
\textit{Spend\,Electronics} & .04 & .11 & .16 & .20 & .28 
\end{tabular}
\hfill
\begin{tabular}{l|ccccc}
\textbf{5 Weeks}& \multicolumn{5}{c}{\textit{depth in tree}}\\
 & $1$ & $\leq 2$ & $\leq 3$ & $\leq 4$ & $\leq 5$ \\
\hline
\rule{0pt}{1.25\normalbaselineskip}\textit{SI\,Fashion}
& .38 & .42 & .42 & .42 & .42\\
\textit{Bids\,Other} & .26 & .52 & .55 & .57 & .67\\
\textit{VI\,sessions} & .02 & .08 & .20 & .26 & .36\\
\textit{lastmonth} & .03 & .13 & .31 & .59 & .72\\
\textit{SRP\,views} & .00 & .06 & .11 & .22 & .42\\
\textit{VI\,views} & .01 & .03 & .08 & .14 & .27\\
\textit{SYI\,views} & .00 & .02 & .04 & .09 & .13
\end{tabular}}

\vskip .25cm
\captionof{table}{\label{tab:trees} Posterior probabilities of the TOT (CART for $y^\star$)  algorithm splitting at or above depths 1 to 5 on each of the variables in the corresponding sample TOT tree (see Figure \ref{fig:trees}), after  one and five weeks of experimentation.  These probabilities are obtained from a Bayesian Forest (sample) of 1000 TOT trees fit under the same settings as our sample trees: maximum depth of 5 and minimum leaf size of 100,000.}
\end{figure}

\subsubsection{Example: posterior uncertainty for transformed outcome trees}

Sample-fit TOT trees for our EBay example experiment are shown in Figure
\ref{fig:trees}, fit to the data accumulated through one and five weeks of
experimentation.  The leaf nodes are marked with the
corresponding prediction rule: $\bar y _{\sf s}^\star$, the mean of
$y^\star_i$ over $i \in \sf s$, which is an estimate for
$\ds{E}_d[\upsilon_i(1) -
\upsilon_i(0)\mid i \in {\sf s}]$ following (\ref{eystar}).  
Recall that these are dollar-value effects.
We fit all of our trees and forests via adaptations of MLLib's decision tree
methods in Apache Spark, and in this case the algorithm stops at either a
maximum depth of 5 or a minimum leaf size of 100,000 users.
\cite{athey_machine_2015} recommend the use of cross-validated pruning for TOT
tree fitting.  In our examples, cross-validation selects deeper trees than
those shown in Figure \ref{fig:trees}, such that these can be viewed as the
\textit{trunks} of some more complex optimal sample TOT.

The population version of the TOT algorithm simply replaces $y_i$ with
$y^\star_i$ in (\ref{impurity}), and a posterior sample over TOT trees -- a
Bayesian Forest -- is obtained via Bayesian bootstrapping as described above.
We fit Bayesian Forests of 1000 trees to study the uncertainty associated
with the TOT trees in Figure \ref{fig:trees}.  For the
variables split upon in each sample TOT tree, Table
\ref{tab:trees} contains the posterior probability that  each variable is
split upon, at or above a given depth, for a new realization of the DGP.  This
is simply  the proportion of trees in the forest in which such splits occur.
The internal decision nodes in Figure \ref{fig:trees}
are colored according to these probabilities.

After one week and 7.45 million users, only
\textit{Bids\,Fashion} -- the number of bids on `fashion' items -- is split upon with greater than 1/3 probability at a depth $\leq 5$.
After observing 5 weeks of purchasing from 13.22 million users, the structure
is more stable:  five variables occur in more than 1/3 of
depth-5 trees.  Two variables occur with probability greater than
1/2: the \textit{lastmonth} indicator, for whether the user made a purchase in
the past month; and
\textit{Bids\,Other}, the number of bids on un-categorized items (the split
location for \textit{Bids\,Other} was always between 6 and 10). We 
could hence, say, partition users into four groups according to the splits on
\textit{lastmonth}  and \textit{Bids\,Other} and have a better than 1/2 chance
that for any posterior DGP realization a similar partitioning would be
included in the top of the corresponding TOT fit.

However, even after 5 weeks,
there remains considerable uncertainty associated with the full tree structure. For example, the very first (root) split in the sample tree occurs in only 40\% of depth-5 trees, it targets a small subset of the data (less than 1\% of users had \textit{SI\,Fashion} $>15$), and it predicts an extreme treatment effect for this subset (-\$38.77 as the effect of slightly larger images).  Moreover, at a depth of 5 all variables except for \textit{lastpurch} have low split probability.  This uncertainty contradicts the examples and results in \cite{taddy_forests_2015}, which finds high posterior probability for the trunks of CART fits and takes advantage of this stability for efficient computation.  We hypothesize that the difference here is due to the tiny signal available for prediction of HTE in our (and probably many other) digital experiments.

\subsection{Bayesian Forest HTE prediction}

A single CART fit is a fragile object; even if the trunks are more stable than
we find in the example above, deep tree structure will have near-zero
posterior probability.  Splits that cross-validated pruning finds useful for
out-of-sample prediction will often disappear under small jitter to the
dataset.  See \citet{breiman_heuristics_1996} for the classic study of this
phenomenon, which was the motivation for his Random Forest algorithm.  Breiman
showed that by averaging across many trees, each individually unstable and
over-fit, he could obtain a response surface that was both stable  and a
strong performer in out-of-sample prediction.  The act of averaging removes
noisy structure that exists in only a small number of trees, and it smooths
across uncertainty, e.g., about split locations or the order of nodes in a
tree path.

From our perspective, a Bayesian Forest (which is nearly equivalent to a
Random Forest) is a posterior over CART predictors.  The average leaf
value associated with a given $\bm{x}$ is the \textit{posterior mean
prediction rule}. This contrasts with the predictions implied by the single
sample CART tree, which is the CART \textit{prediction rule at posterior mean DGP}, where
$\bs{\theta}=\bm{1}$. Experience shows that this can make a big
difference: the forest average response surface will be different from and provide better
prediction than the sample CART
tree. (More generally, see \cite{clyde_bagging_2001} for discussion on the Bayesian bootstrap and model averaging.)  


For our final example, we consider the posterior distribution on the difference
between  {\it  two} prediction rules: CART fit to each of treatment and control DGPs. 
Write $\hat y_{\sf
d}(\bm{x})$ for the prediction rule at $\bm{x}$ resulting from population CART
fit to support $\bm{Z}_{\sf d} = \{ \bm{x}_{i},y_i: i
\in {\sf d}\}$ with weights $\bs{\theta}_{\sf d}$.  That is, 
if the realized CART fit allocates $\bm{x}$ to the leaf node containing observations in set $\bm{\sf s}$, then $\hat y_{\sf
d}(\bm{x}) = \mu_\textsf{s} = \bm{y}_\textsf{s}'\bs{\theta}_\textsf{s}/|\bs{\theta}_\textsf{s}|$ as described in (\ref{sse}).
Thus 
$\hat y_{\sf
d}(\bm{x})$ is a random variable and so is the predicted treatment effect
\begin{equation}\label{treetreat}
\hat y_{\sf
t}(\bm{x}) - \hat y_{\sf
c}(\bm{x}).
\end{equation}
As in Section \ref{sec:linear}, the DGPs for  treatment and control are independent from each other and we can obtain posterior samples of (\ref{treetreat}) via separate Bayesian Forests for each treatment group.

The framework implied by (\ref{treetreat}) is related to a
semi-parametric literature \citep[e.g.][]{hill_bayesian_2011,green2012modeling,grimmer_estimating_2013,imai2013estimating} that studies the difference between
flexible regression functions in each of the treatment groups.  
In a prominent example, \cite{hill_bayesian_2011}  applies Bayesian additive regression trees
\citep[BART;][]{chipman_bart:_2010} and interprets the difference
 between   posterior predictive distributions across treatment groups as
 effect heterogeneity. In contrast to our approach, where the trees are just a
 convenient prediction rule and we do not assume that the data were actually
 generated from a tree, Hill assumes that her regression
 functions are representative of the true underlying DGP.   Which strategy is best will depend upon your application. For example, BART includes a
 homogeneous Gaussian additive error and is thus inappropriate for the
 heteroscedastic errors in Internet transaction data.  BART is outperformed by
 forest algorithms in such settings \citep[see, e.g.,][]{taddy_forests_2015}, but will outperform the forests when the
 homoscedasticity assumption is more valid.

 Due to the similarity between Random and
 Bayesian Forests, our approach is also related to recent work by
 \citet{wager_estimation_2015} on the use of  Random Forests in
 HTE estimation.  Wager and Athey use the forests  to
 construct  confidence intervals for a true treatment effect surface.  This is more
 ambitious than our contribution, which interprets the forest as a
 posterior distribution for optimal prediction of treatment effects within a
 certain class of algorithms.  Indeed, Wager and Athey are studying the frequentist
 properties of our posterior mean.

\begin{figure}[t]
\includegraphics[width=.25\textwidth]{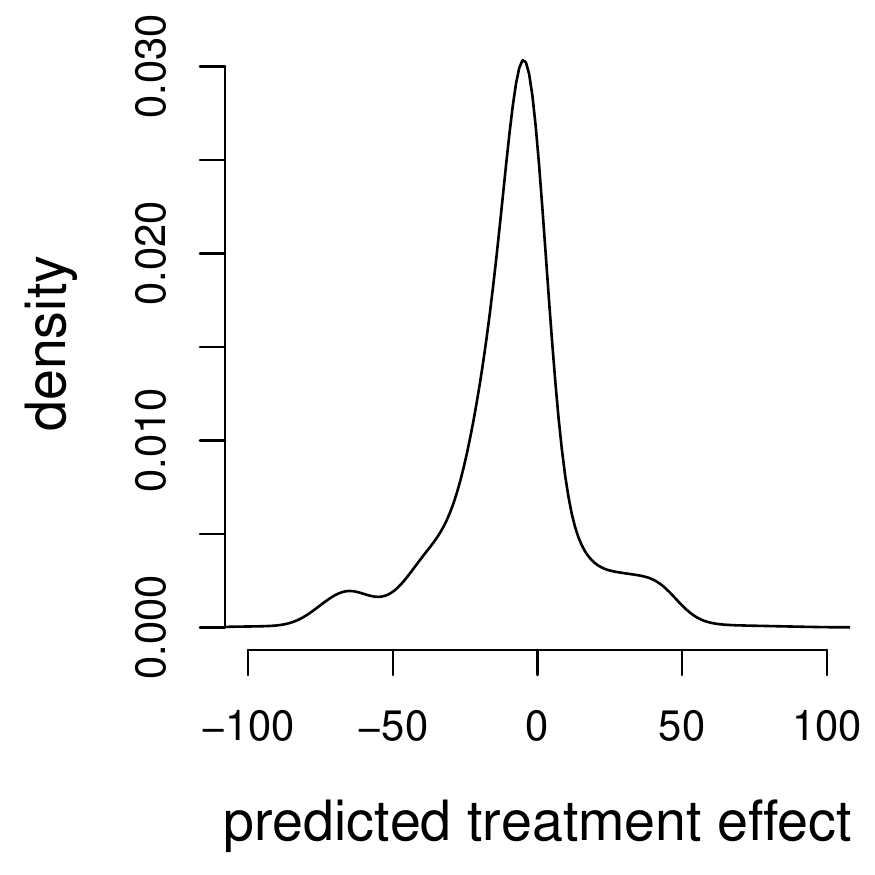}\!
\includegraphics[width=.25\textwidth]{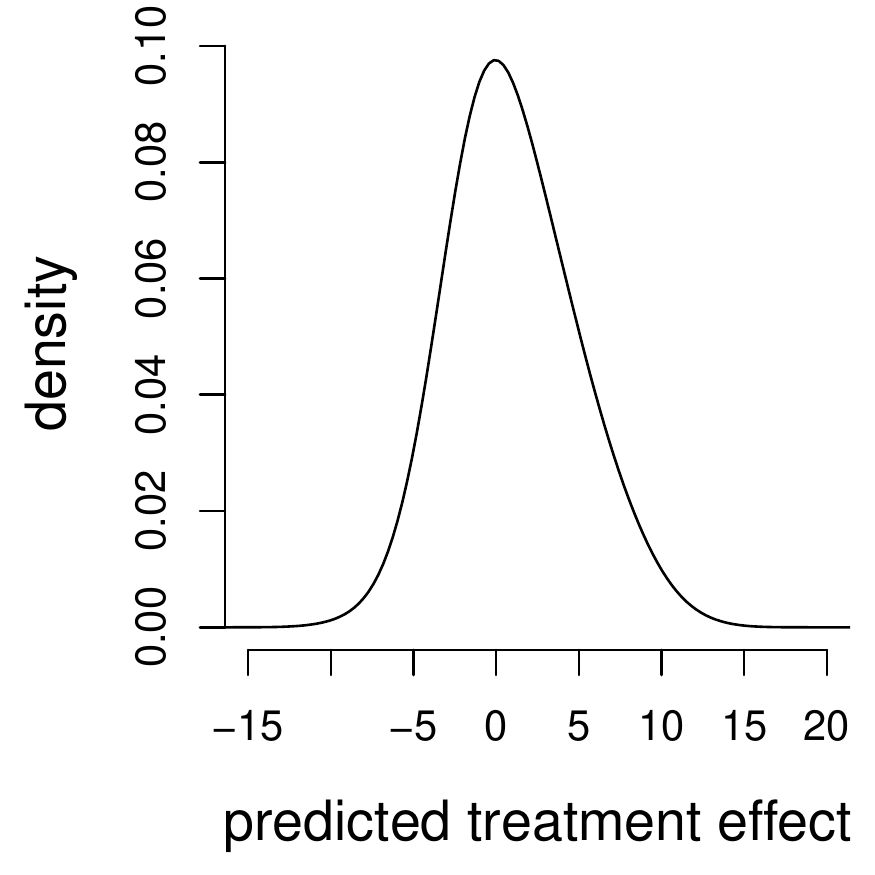}\!
\includegraphics[width=.25\textwidth]{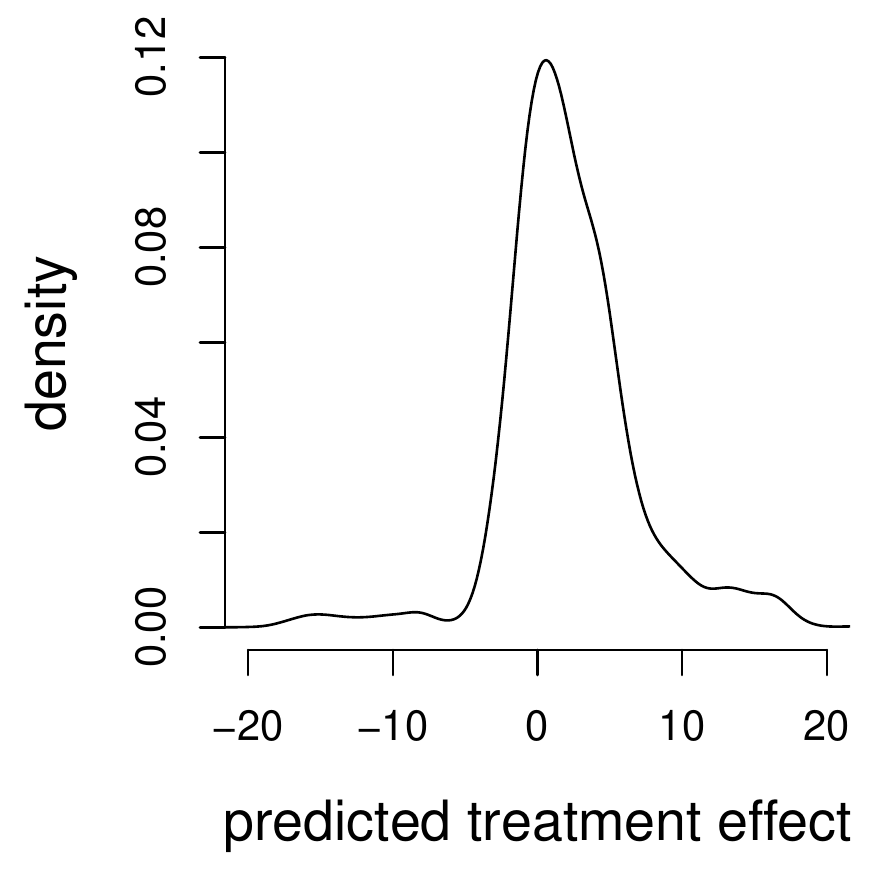}\!
\includegraphics[width=.25\textwidth]{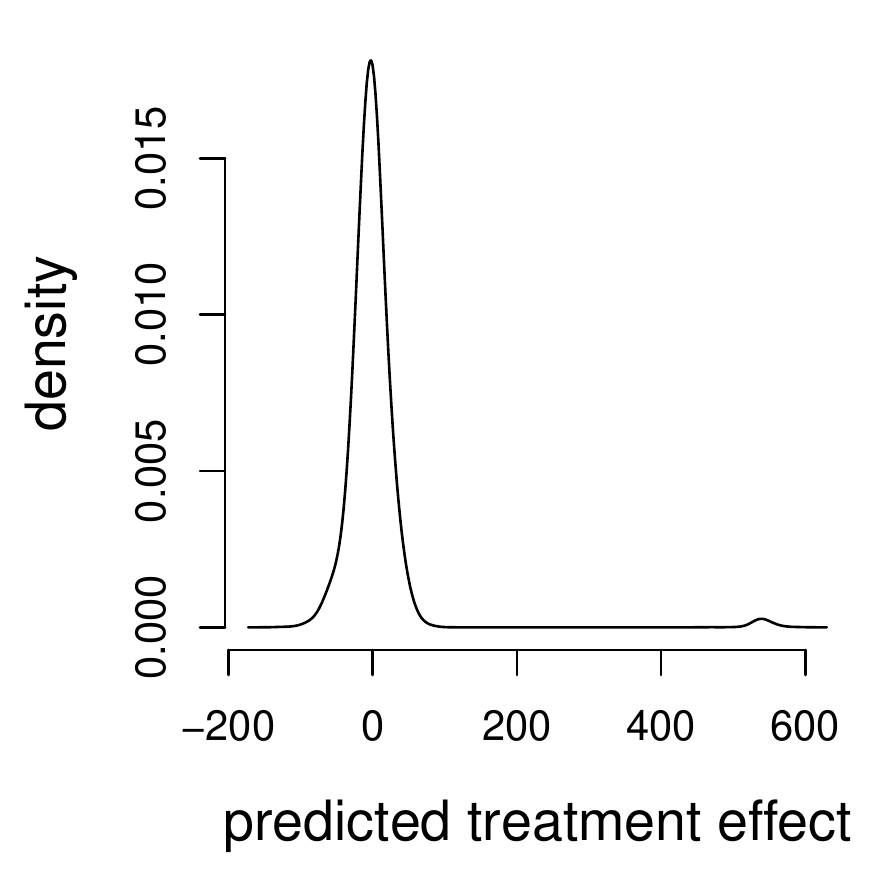}\!
\caption{\label{fig:userposts} Bayesian Forest treatment effects after 5 weeks.  Each plot shows the posterior distribution for population CART treatment effect prediction, $\hat y_{\sf t}(\bm{x}_i) -
\hat y_{\sf t}(\bm{x}_i)$, at  features $\bm{x}_i$ from a sampled user. }
\end{figure}

\subsubsection{Example: Differenced treatment group forests}

Returning to our EBay example, we focus on HTE prediction for the completed
experiment including 13.22 million users over 5 weeks.  Bayesian Forests of
1000 trees each were fit to the treatment and control group samples.  Each
forest is the posterior distribution over a population-CART algorithm run with
maximum depth of 10 and no minimum leaf size. CART was applied without any
random variable subsetting; hence, variability in the resulting prediction
surface is  due entirely to posterior uncertainty about the DGP.

\begin{figure}[t]
\includegraphics[width=\textwidth]{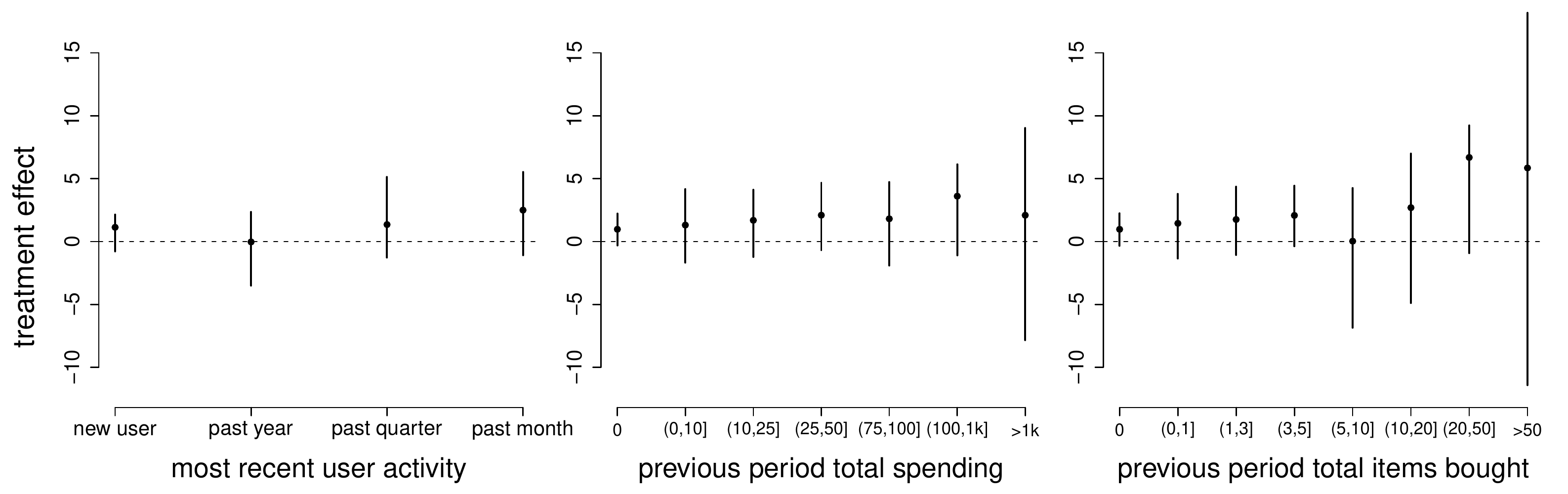}
\caption{\label{fig:maineff} Variable average effects as defined in (\ref{maineff}) for three important user features: the user's last purchase date and their spending (in dollars and items bought) over the period prior to this experiment.  Points mark the posterior mean and line segments extend from the $10^{th}$ to $90^{th}$ posterior percentile.  }
\end{figure}

The outcome is a posterior sample over prediction rules for the conditional
average treatment effects, $\hat y_{\sf t}(\bm{x}) -
\hat y_{\sf c}(\bm{x})$ as in (\ref{treetreat}).  Figure \ref{fig:userposts} shows
four example posteriors for individual user treatment effects;
this type of uncertainty quantification is available for any new user
whose treatment effect you wish to predict.  It is also possible to summarize, for a given DGP realization, the average treatment effect conditional on variable $j$ being in set $\mc{X}$ as
\begin{equation}\label{maineff}
\hat y_{\sf t}^j(\mc{X}) -
\hat y_{\sf c}^j(\mc{X}) := \frac{\sum_{i: x_{ij} \in \mc{X}} \theta_i \left( \hat y_{\sf t}(\bm{x}_i) -
\hat y_{\sf c}(\bm{x}_i)\right)}{\sum_{i: x_{ij} \in \mc{X}} \theta_i}.
\end{equation}
The sum in (\ref{maineff}) is over all observations in the sample (both treatment and control groups) that have their $j^{th}$ feature in $\mc{X}$.
Figure \ref{fig:maineff} shows change in the posterior distributions for conditional average treatment effects corresponding to change in the user's last purchase date and their spending (in dollars and items bought) over the period prior to the experiment. Both posterior mean and uncertainty tend to increase for groups of more active users.   As in our OLS analysis of Figure \ref{fig:gammapost}, the posteriors can be highly skewed.

Finally, each DGP realization provides a prediction for the average treatment effect,
\begin{equation}\label{treeate}
\hat y_{\sf t} -
\hat y_{\sf c} := \frac{1}{|\bs{\theta}|}\sum_{i=1}^n \theta_i \left( \hat y_{\sf t}(\bm{x}_i) -
\hat y_{\sf c}(\bm{x}_i)\right).
\end{equation}
The random $\bs{\theta}$ play a role here both in weighting each treatment effect prediction, $\hat y_{\sf t}(\bm{x}_i) -
\hat y_{\sf c}(\bm{x}_i)$, and in the CART fits that underly those predictions.  The last row of Table \ref{tab:perweek} shows posterior mean and standard deviation for $\hat y_{\sf t} -
\hat y_{\sf c}$ from (\ref{treeate}) after 1-5 weeks of experimentation.  We have no basis here to argue that  this statistic  is preferable to unadjusted $\mu_{\sf t} - \mu_{\sf c}$ or the adjusted $\bs{\mu}_{\bm{x}}'(\bs{\beta}_{\sf t} - \bs{\beta}_{\sf c})$ metrics of Section \ref{sec:linear}.2; however, it presents an intuitively appealing option if you believe that the expected response within each treatment group is nonlinear in $\bm{x}$.

\section{Discussion}
\label{sec:discuss}

This article outlines a nonparametric Bayesian framework for treatment effect analysis in A/B experiments. The approach is simple, practical, and scalable.  It applies beyond the two studied classes of HTE statistics; for example, an earlier version of the work \citep{taddy2014heterogeneous} considered HTE summarization via moment conditions and our CART trees are just one possible prediction rule amongst  many available machine learning tools.

One area for future research is in semi-parametric extensions of this
framework.  For example, we know from existing theory on the frequentist
nonparametric bootstrap that it can fail for distributions with infinite
variance \citep{athreya1987bootstrap}. This scenario could occur in digital experiments where the response is
extremely heavy tailed.  In response, \cite{taddy_heavytails_2015} propose
combining the Dirichlet-multinomial model with a parametric tail distribution.

Throughout, we have referenced  large existing literatures on Bayesian
parametric and semi-parametric and frequentist analysis of HTE.  We are not
aiming to replace these existing frameworks, nor are we advocating for any one
HTE statistic over another. Instead, we simply present a novel set of Bayesian
nonparametric analyses for some common and useful tools.  The hope is that frequentists
and parametric Bayesians alike will benefit from this alternative point of
view.

\bibliographystyle{chicago}
\bibliography{taddy}

\appendix

\section{Population OLS gradient}
\label{sec:olsgrad}

Define
 $\bm{S}_{\sf d} = \bm{X}_{\sf d}'\bs{\Theta}_{\sf d}\bm{X}_{\sf d}$ and use $\nabla \bm{v}$ to denote the $\mr{len}(\bm{v}) \times n_{\sf d}$ gradient of $\bm{v}$ on $\bs{\theta}_{\sf d}$.
Then 
\begin{align}
\nabla \bs{\beta}_{\sf d} &= \nabla\!\left(\bm{S}^{-1}_{\sf d}\bm{X}_{\sf d}'\bs{\Theta}_{\sf d}\bm{y}_{\sf d}\right) \label{betagrad} 
 \\
 &=  
\bm{S}^{-1}_{\sf d}
\nabla\mr{vec}(\bm{X}_{\sf d}'\bs{\Theta}_{\sf d}\bm{y}_{\sf d}) + (\bm{y}_{\sf d}'\bs{\Theta}_{\sf d}\bm{X}_{\sf d}\otimes \bm{I}_{p})\nabla\mr{vec}(\bm{S}^{-1}_{\sf d})
\notag\\
&= \bm{S}^{-1}_{\sf d}(\bm{y}_{\sf d}'\otimes\bm{X}_{\sf d}')\nabla\mr{vec}(\bs{\Theta}_{\sf d})
-(\bm{y}_{\sf d}'\bs{\Theta}_{\sf d}\bm{X}_{\sf d}\otimes \bm{I}_{p})(\bm{S}_{\sf d}^{-1}\otimes\bm{S}_{\sf d}^{-1})(\bm{X}_{\sf d}'\otimes\bm{X}_{\sf d}')\nabla\mr{vec}(\bs{\Theta}_{\sf d})
\notag \\
& = \bm{y}_{\sf d}'\otimes\bm{S}^{-1}_{\sf d}\bm{X}_{\sf d}' 
- \bm{y}_{\sf d}'\bs{\Theta}_{\sf d}\bm{X}_{\sf d}\bm{S}_{\sf d}^{-1}\bm{X}_{\sf d}'
\otimes\bm{S}_{\sf d}^{-1}\bm{X}_{\sf d}'\nabla\mr{vec}(\bs{\Theta}_{\sf d})\notag\\
& = (\bm{y}_{\sf d} - \bm{X}_{\sf d}\bs{\beta}_{\sf d})'\otimes\bm{S}_{\sf d}^{-1}\bm{X}_{\sf d}'\nabla\mr{vec}(\bs{\Theta}_{\sf d})\notag
\end{align}
via repeated applications of $\mr{vec}(\bm{ABC}) = (\bm{C}'\otimes\bm{A})\mr{vec}(\bm{B})$ and $(\bm{A}\otimes\bm{B})(\bm{C}\otimes\bm{D}) = \bm{AC}\otimes\bm{BD}$ for appropriately sized matrices, and using the chain rule with a standard result from matrix calculus to get $\nabla \bm{S}_{\sf d}^{-1} = \partial \bm{S}_{\sf d}^{-1}/ \partial \bm{S}_{\sf d}\nabla \bm{S}_{\sf d} = -(\bm{S}_{\sf d}^{-1}\otimes\bm{S}_{\sf d}^{-1})\nabla \bm{S}_{\sf d}$.  Since $\nabla \mr{vec}(\bs{\Theta}_{\sf d}) = \bm{1}_{n_{\sf d}^2}$, the formula in (\ref{betagrad}) reduces to 
$\nabla \bs{\beta}_{\sf d} = \bm{S}_{\sf d}^{-1} \bm{X}_{\sf d}\mr{diag}(\bm{y}_{\sf d}-\bm{X}_{\sf d}\bs{\beta}_{\sf d})$.

\section{Posterior inference for regression-adjusted ATE}
\label{sec:regappendix}

Our first-order approximation to $\bs{\mu}_{\bm{x}}'\left(\bs{\beta}_{\sf t} -
\bs{\beta}_{\sf c}\right)$ is  $\bm{\bar
x}'(\bs{\tilde\beta}_{\sf t} -
\bs{\tilde \beta}_{\sf c})$. Writing $\mr{var}(\bs{\tilde \beta}_{\sf d}) = \bs{\Sigma}_{\bs{\tilde \beta}_{\sf d}} $, this approximation has variance $\bm{\bar x}' \left(\bs{\Sigma_{\tilde \beta_{{\sf t}} }}+
\bs{\Sigma_{\tilde \beta_{{\sf c}} }}\right)\bm{\bar x}$.
\begin{theorem}
\begin{align}
\mr{var}\left(\bm{\bar x}'\left[\bs{\tilde\beta}_{\sf t} -
\bs{\tilde \beta}_{\sf c}\right]\right) &=  
\frac{s^2_{\bm{y}_{\sf c}}}{n^2_{\sf t}} +  
\frac{s^2_{\bm{y}_{\sf c}}}{n^2_{\sf c}}  - \left( \frac{R^2_{\sf t}s^2_{\bm{y}_{\sf c}}}{n^2_{\sf t}} +  
\frac{R^2_{\sf c}s^2_{\bm{y}_{\sf c}}}{n^2_{\sf c}} \right) \notag\\ &+
(\bm{\bar x}-\bm{\bar x}_{\sf t})' \bs{\Sigma_{\tilde \beta_{{\sf t}} }}(\bm{\bar x}-\bm{\bar x}_{\sf t}) +
(\bm{\bar x}-\bm{\bar x}_{\sf c})' \bs{\Sigma_{\tilde \beta_{{\sf c}} }}(\bm{\bar x}-\bm{\bar x}_{\sf c}),\label{ratevar}
\end{align}\label{ratevarthm} where $s^2_{\bm{v}} = \bm{v}'\bm{v} - n_v
\bar{\bm{v}}^2$ for generic length-$n_v$ vector $\bm{v}$ and $R^2_d = 1 -
s^2_{\bm{r}_{\sf d}}/s^2_{\bm{y}_{\sf d}}$.
\end{theorem}
\noindent
\begin{proof}
Consider the shifted OLS projections $\bs{\dot \beta}_{\sf d}
= (\bm{\dot X}_{\sf d}'\bs{\Theta}_{\sf d}\bm{\dot X}_{\sf d})^{-1}\bm{\dot X}_{\sf d}\bs{\Theta}_{\sf d}\bm{y}_{\sf
d}$, using design matrix $\bm{\dot X}_{\sf d}$ that has been centered within each
group (except for the intercept) so that $\bm{\dot X}_{ \sf d}'\bs{\theta}_{ \sf d} =
\left[\begin{smallmatrix}n_{\sf d}\\ \bm{0}_{p-1}\end{smallmatrix}\right]$.  Say $\bs{\tilde{\dot\beta}}_{\sf d}$ is the first-order approximation of (\ref{taylorols}) applied to $\bs{\dot{\beta}}_{\sf d}$, with variance $\mr{var}(\bs{\tilde{\dot\beta}}_{\sf d}) = 
(\bm{\dot X}^{\prime}_{\sf d}\bm{\dot X}_{\sf d})^{-1}\bm{\dot X}_{\sf d}^{\prime}\bm{R}_d\bm{R}_d\bm{\dot X}_{\sf d}(\bm{\dot X}_{\sf d}^{\prime}\bm{\dot  X}_{\sf d})^{-1}$.  Note that the residuals $\bm{r}_{\sf d}$ are unchanged and that the non-intercept coefficients are exactly equal: $\tilde{\dot\beta}_{{\sf d}j} = \tilde{\beta}_{{\sf d}j}$ for $j>1$.  Thus
$\bm{\bar
x}'\bs{\tilde\beta}_{\sf d}  = \tilde{\dot\beta}_{ {\sf d}1}  + [\bm{\bar x}-\bm{\bar x}_{\sf
d}]'\bs{\tilde{\dot\beta}}_{ {\sf d}}  =  \tilde{\dot\beta}_{ {\sf d}1} + [\bm{\bar x}-\bm{\bar x}_{\sf
d}]'\bs{\tilde \beta}_{ {\sf d}}$
with variance 
$\mr{var}( \bs{\tilde{\dot\beta}}_{ {\sf d}1} + [\bm{\bar
x}-\bm{\bar x}_{\sf d}]'\bs{\tilde\beta}_{\sf d}) 
 = \frac{1}{n_{\sf d}^2}\bm{r_{\sf d}}'\bm{r_{\sf d}} + (\bm{\bar x}-\bm{\bar x}_{\sf d})' \bs{\Sigma_{\tilde \beta_{{\sf d}} }}(\bm{\bar x}-\bm{\bar x}_{\sf d})$.  Using $\bm{r_{\sf d}}'\bm{r_{\sf d}} = (1-R_{\sf d}^2)s^2_{\bm{y}_{\sf d}}$ and summing 
  $\mr{var}(\bm{\bar x}'\bs{\tilde\beta}_{\sf t}) +\mr{var}(\bm{\bar x}\bs{\tilde\beta}_{\sf c})$ completes the result.
\end{proof}
\noindent 
Making the rough equivalences $n_{\sf d} \approx n_{\sf d} + 1$ and  $\bm{\bar
x}-\bm{\bar x}_{\sf c} \approx \bm{0}$, the result in (\ref{ratevar}) leads to
our expression in (\ref{roughatevar}). Note that (\ref{ratevar}) ignores
variance in the covariate mean, $\bs{\mu}_{\bm{x}} =
\bm{X}'\bs{\theta}/|\bs{\theta}|$, which is correlated with
$\bs{\tilde\beta}_{\sf t} -
\bs{\tilde \beta}_{\sf c}$ and has variance
$
= \frac{1}{n+1}\left[ \frac{1}{n}\bm{X}'\bm{X} 
 -  \bm{\bar x} \bm{\bar x}' \right].$

\end{document}